\documentclass[amsmath, amssymb, english, aps, prb, twocolumn, reprint,floatfix, superscriptaddress,showpacs]{revtex4-1}

\usepackage[english]{babel}
\usepackage{graphics}
\usepackage{graphicx}
\usepackage{epsfig}
\usepackage{verbatim}
\usepackage[usenames,dvipsnames]{color}

\begin{document}

\title{Low-energy physics of three-orbital impurity model with Kanamori interaction}
\author{Alen~Horvat}
\affiliation{Jo\v{z}ef Stefan Institute, Jamova~39, Ljubljana, Slovenia}
\author{Rok~\v{Z}itko}
\affiliation{Jo\v{z}ef Stefan Institute, Jamova~39, Ljubljana, Slovenia}
\affiliation{Faculty of Mathematics and Physics, University of
Ljubljana, Jadranska 19, Ljubljana, Slovenia}
\author{Jernej~Mravlje}
\affiliation{Jo\v{z}ef Stefan Institute, Jamova~39, Ljubljana, Slovenia}

\begin{abstract}
We discuss the low-energy physics of the three-orbital Anderson impurity model
with the Coulomb interaction term of the Kanamori form which has orbital SO(3)
and spin SU(2) symmetry and describes systems with partially occupied $t_{2g}$
shells. We focus on the case with two electrons in the impurity that is relevant
to Hund's metals. Using the Schrieffer-Wolff transformation we derive
an effective
Kondo model with couplings between the bulk and impurity electrons expressed in
terms of spin, orbital, and orbital quadrupole operators. The bare spin-spin
Kondo interaction is much smaller than the orbit-orbit and spin-orbital
couplings or is even ferromagnetic. Furthermore, the perturbative scaling
equations indicate faster renormalization of the couplings related to orbital
degrees of freedom compared to spin degrees of freedom. Both mechanisms lead to
a slow screening of the local spin moment. The model thus behaves similarly to
the related quantum impurity problem with a larger SU(3) orbital symmetry
(Dworin-Narath interaction) where this was first observed. We find that the two
problems actually describe the same low-energy physics since the SU(3) symmetry
is dynamically established through the renormalization of the
splittings of coupling constants to
zero. The perturbative renormalization group results are corroborated with the
numerical-renormalization group (NRG) calculations. The dependence of spin Kondo
temperatures and orbital Kondo temperatures as a function of interaction
parameters, the hybridization, and the impurity occupancy is calculated and
discussed.

\end{abstract}

\maketitle

\def\mTk{T_{\mathrm{K}}}
\def\Tk{$T_{\mathrm{K}}$}
\def\Tksp{$T_K^{\mathrm{spin}}$}
\def\Tkorb{$T_K^{\mathrm{orb}}$}
\def\mTksp{T_K^{\mathrm{spin}}}
\def\mTkorb{T_K^{\mathrm{orb}}}
\section{Introduction}

The theoretical work of recent years has led to a considerably better
understanding of the origin of electronic correlations in materials with wide
bands and relatively weak Coulomb interactions, such as iron-based
superconductors and ruthenates.  Based on the dynamical mean-field theory
calculations (DMFT)~\cite{georges_review_dmft} it has been realized  that a
small multiplet splitting coming from the Hund's rule part of the Coulomb
interaction ($J\ll U<W$, with $W$ the bandwidth, $U$ Hubbard interaction) has
drastic effects at low energy
scales~\cite{haule09,werner08,mravlje11,Hansmann_localmoment_prl,Georges2013}.
This has important consequences for the physics of these materials that are
hence being referred to as the Hund's
metals~\cite{Yin_kinetic_frustration_allFeSC, Georges2013,
Medici2015,Bascones2015}.

Impurity models play a major role in the DMFT studies since the problem of the
bulk is mapped to a problem of a quantum impurity embedded in a
self-consistently determined bath.  It is interesting to note that whereas in
the single-orbital setting the relevant impurity problem was well
explored~\cite{hewson_book_1993} prior to the development of the DMFT, this is not
the case for multi-orbital systems where the DMFT calculations
preceded~\cite{haule09,werner08,mravlje11,Hansmann_localmoment_prl,Georges2013}
the detailed investigation of the impurity models upon which those calculations
are based. The discovery of the strong influence of the Hund's rule coupling within the
DMFT has encouraged studies of multi-orbital effects also for adatoms on
metal surfaces.~\cite{Khajetoorians2015,Dang2015c}

In this paper we study the three-orbital impurity problem with Kanamori interaction
\begin{equation}
	\label{eq:Himp0}
	H_{\mathrm{imp}} = \frac{1}{2}(U - 3J)N_{d}(N_{d}-1)
	-2J \mathbf{S}^{2} -
	\frac{J}{2} \mathbf{L}^{2},
\end{equation}
relevant for instance to the DMFT description of a transition-metal
oxide with partially occupied $t_{2g}$ shells. In three-orbital
systems, the physics of Hund's metals occurs at occupancy
$N_{d}=2$~\cite{Medici2011}. $\mathbf{L}, \mathbf{S}$ are the orbital
momentum and spin operators, respectively.  Hamiltonian in
Eq.~\eqref{eq:Himp0} has a SU(2) spin and SO(3) orbital symmetry.  The
low-energy properties of the model defined by Eq.~\eqref{eq:Himp0}
have not been studied so far.

The effects of the Hund's rule coupling were explored for several simpler
(mostly two-orbital) models~\cite{schrieffer_japplphys_1967,jayaprakash81,
jones87, Kuramoto1998,
kusunose97,yotsuhashi01,pruschke_epjb_2005,nevidomskiy09,Nishikawa:2012gf}.
The common conclusion
of these works is that the Hund's rule coupling suppresses the Kondo temperature
through reduced exchange coupling of the low-lying impurity spin
degrees of freedom with conduction
electrons.

More recently, a  Dworin-Narath (DN) impurity model~\cite{Dworin1970} was
studied~\cite{yin2012,Aron2015,Stadler2015}.  The DN model is described in terms
of the simplified interaction Hamiltonian
\begin{equation}
	\label{eq:Himp1}
	H_{\mathrm{imp}} =  \frac{1}{2}\left(U - 3J\right)N_{d}(N_{d}-1)
	-2J \mathbf{S}^{2},
\end{equation}
which is similar to Eq.~\eqref{eq:Himp0}, but without the orbital part
of the Hund's interaction, $ -(J/2) \mathbf{L}^2$. The DN model has a
higher SU(3) orbital symmetry and different fixed points.
This work has led to important qualitative insights into the physics of
Hund's metal. Namely, Refs.~\cite{yin2012,Aron2015} derived a Kondo Hamiltonian
with a SU($M$) orbital and SU($N$) spin symmetry and argued that the
key property is that the
spin-spin Kondo coupling is ferromagnetic (or small) and that a two-stage
screening of spin and orbital degrees of freedom occurs (see also an earlier
pioneering study~\cite{okada73}). In Ref.~\cite{Aron2015} a
renormalization group (RG) analysis stressed the importance of different
spin and orbital degeneracy. These findings were corroborated by the
numerical-renormalization-group (NRG) study in Ref.~\cite{Stadler2015}.

Given the deep implications of these results it is important to
investigate the problem for the more realistic interaction term that
is actually used in the DMFT calculations.  In this paper we
investigate the low-energy physics of the Anderson impurity model
(AIM) with Kanamori interaction at occupancy close to $N_d=2$ which is
relevant to Hund's metals.  We derive the corresponding Kondo
Hamiltonian using the Schrieffer-Wolff transformation.
The distinction between the Kanamori and the Dworin-Narath Hamiltonian is found to become
asymptotically irrelevant: at low energies, the orbital SO(3)
symmetry is dynamically enlarged to the larger SU(3) symmetry.
Consequently, the qualitative picture of the two-stage screening applies also for
the Kanamori Hamiltonian.
We also performed the NRG simulations that confirm these weak-coupling RG
findings.  We calculated the dependence of spin and orbital Kondo
temperatures for a range of parameters and electron occupancies.
Except at very low values of the Hund's rule coupling strength, the
spin Kondo temperature is significantly smaller (an order of magnitude
or more). The smaller bare value of the spin-Kondo coupling as well as
its slower running both contribute to such behavior.

The paper is structured as follows. In Sec.~\ref{sec:model} we start with the
description of the model. In Sec.~\ref{sec:sw} we present the Schrieffer-Wolff
transformation, the resulting Kondo Hamiltonian, and the Kondo couplings. In
Sec.~\ref{sec:poor} we discuss the RG flow using the poor man's scaling
approach. In Sec.~\ref{sec:nrg} we give the NRG results.  In Sec.~\ref{sec:conc}
we conclude with a discussion of the implications of our results and
with prospects
for future work. In appendices~\ref{app:kondo_ham} and~\ref{app:rg} we give
technical details on the derivation of Kondo Hamiltonian and RG flow,
respectively.
In appendix~\ref{app:symmetric} we express the Kondo Hamiltonian
in terms of rescaled couplings in a way that the couplings and the scaling
equations are equal when the Hund's coupling is zero.
In appendix~\ref{app:kan_vs_dn} we compare the behavior of
Dworin-Narath and Kanamori models.

\section{Impurity model}
\label{sec:model}

The  impurity models of interest to this paper  can be written in the following way:
\begin{eqnarray}
    H_{\mathrm{bath}} &=& \sum_{k,m,\sigma} \epsilon_{k}
    c^\dag_{km\sigma} c_{km\sigma},
    \label{eq:Hyb} \\
    H_{\mathrm{hyb}} &=& \sum_{k,m, \sigma} V_{k} c_{km \sigma}^{\dagger}d_{m \sigma}
    + \mathrm{h.c.} \nonumber\\
    &=& V\sum_{m, \sigma} c_{m \sigma}^{\dagger}d_{m\sigma} + \mathrm{h.c.},\\
    \label{eq:imp-general}
    H_{\mathrm{imp}} &=&-2J \mathbf{S}^{2} -\alpha\frac{J}{2} \mathbf{L}^{2}
    \nonumber\\
    &\quad& +\frac{U - 3J }{2}N_{d}(N_{d}-1) + \epsilon_0 N_d,
\end{eqnarray}
with
\begin{equation}
\begin{split}
N_d &= \sum_{m,\sigma} d^\dag_{m\sigma}d_{m\sigma}, \\
\mathbf{S} &= \sum_{m} d^\dag_{m\sigma} \left( \frac{1}{2}
\boldsymbol{\sigma}_{\sigma\sigma'} \right) d_{m\sigma'}, \\
\mathbf{L} &= \sum_\sigma d^\dag_{m\sigma} \mathcal{L}_{mm'} d_{m'\sigma}.
\end{split}
\end{equation}
The operators $c^{(\dagger)}_{m \sigma}$ and $d^{(\dagger)}_{m
\sigma}$ annihilate (create) bath and impurity electrons with spin
\(\sigma=\pm 1/2\) in orbital $m\in \{1,2,\dots,M\}$, $M$ being the
number of orbitals.  The non-interacting conduction electrons
($H_\mathrm{bath}$) have energy \(\epsilon_{k}\), which corresponds to
a flat density of states $\rho_0=1/2D_0$ with half-bandwidth $D_0$.
In the hybridization function ($H_\mathrm{hyb}$) we
use the notation $\sum_{k} V_{k} c_{km \sigma} = V c_{m \sigma}$. The
hybridization strength is defined as $\Gamma=\pi \rho_0 V^2$.

The interaction of the electrons on the impurity is described by the
term $H_{ \mathrm{imp}}$ where we introduced the parameter \(\alpha\)
that tunes the impurity interaction between Dworin-Narath
(\(\alpha=0\)) and the Kanamori (\(\alpha=1\)) case in a continuous
way. We will refer to the impurity model above as the Anderson
impurity model (AIM) to distinguish it from the Kondo model defined
in the following.
$N_d$ is the total impurity charge operator, $\mathbf{S}$
is the total impurity spin operator ($\boldsymbol{\sigma}$ are
Pauli matrices), and $\mathbf{L}$ is the total impurity orbital
angular momentum ($\mathcal{L}$ are spin-1 matrices for $M=3$).
The spin and orbital momentum operators obey the Lie algebra
commutation relations and are normalized such that: $\mathrm{Tr}(X^{
\alpha} X^{ \beta})=2 \delta_{ \alpha, \beta}, X \in\{L,S\}$.

In the following section we derive an effective Kondo Hamiltonian for
the simplest realistic model that captures the Hund's physics: the
three orbital ($M=3$) AIM with two electrons or holes occupying the
impurity such that the ground state orbital moment and spin are $L=1$,
$S=1$.

We choose units such that $D_0=1$, $k_{B}=1$, $g\mu_{B}=1$.

\section{Kondo Hamiltonian and RG analysis}
\label{sec:sw}

\subsection{Schrieffer-Wolff transformation}

To investigate the low-energy behavior of coupled bath and impurity
electrons we derive an effective Kondo Hamiltonian in which the charge
fluctuations on the impurity are suppressed.
This is achieved using the canonical
Schrieffer-Wolff transformation.~\cite{Schrieffer1966} The interaction term that
is induced by virtual fluctuations from the ground-state impurity multiplet into
the high-energy manifolds with $n\pm 1$ electrons reads
\begin{equation}
    \label{SWtransform}
    H_{\mathrm{K}} = -P_{n}H_{\mathrm{hyb}} \left( \sum_a \frac{P_{n+1}^{a}}{\Delta
    E_{n+1}^{a}} + \sum_b \frac{P_{n-1}^{b}}{\Delta E_{n-1}^{b}}
    \right)H_{\mathrm{hyb}} P_{n}.
\end{equation}
Projector operators $P_{n}$ project onto the atomic ground state multiplet with
valence $n$.  Projectors $P^{a}_{n\pm 1}$ project onto the high energy
multiplets having energy $E^{a}_{n\pm 1}$ (indices $a,b$ denote different
invariant subspaces with respect to $H_{\mathrm{imp}}$) and the virtual
excitation energies are  $\Delta E^{a}_{n\pm 1}= E^{a}_{n\pm 1}-E_{n}$; $E_{n}$
is the ground state energy.

For the case of Kanamori Hamiltonian, Eq.~\eqref{SWtransform} can be rewritten
(see Appendix~\ref{app:kondo_ham} for the derivation) in the following
``Kondo-Kanamori'' form:
\begin{eqnarray}
	\label{Hkk0}
	H_{K} = J_{p} N_{f}+ J_{s}  \mathbf{S \cdot s} + J_{l}  \mathbf{L \cdot l}
	+ J_{q}  \mathbf{Q \cdot q} +\nonumber\\ J_{ls} \mathbf{(L \otimes
	S)\cdot(l\otimes s)}+ J_{qs}
	\mathbf{(Q\otimes S)\cdot(q\otimes s)}.
\end{eqnarray}
$N_f$ is the bulk electron charge operator at the position of the impurity,
$ \mathbf{S}, \mathbf{L}, \mathbf{Q}$ ($ \mathbf{s}, \mathbf{l},
\mathbf{q}$) are total impurity (bath) spin, orbit, and orbital
quadrupole operators, respectively.  The Kondo Hamiltonian contains
besides spin-spin, orbital-orbital and quadrupole-quadrupole
interaction also the mixed spin-orbital and spin-quadrupole products
($ \mathbf{L}\otimes \mathbf{S}, \mathbf{Q}\otimes \mathbf{S}$,
respectively). Eq.~\eqref{Hkk0} can be viewed as a multipole expansion of the
exchange interaction for spin and orbital degrees of freedom; the
highest orders (dipole for spin, quadrupole for orbital momentum) are
related to the degrees of freedom carried by the particles
($\sigma=\pm 1/2$ for spin, $m=1,2,3$ for orbital momentum).
The five (symmetric and
traceless) quadrupole operators are second order orbital tensor operators
defined as
\begin{eqnarray}
	Q^{bc}_{i,j} = \frac{1}{2}\left(L^{b}_{i,m}L^{c}_{m,j}+L^{c}_{i,m}L^{b}_{m,j} \right) - \frac{2}{3} \delta_{b,c} \delta_{i,j},\\
	\mathrm{Tr}(Q^{\alpha}Q^{ \beta}) = 2\delta_{ \alpha, \beta}.
\end{eqnarray}
Quadrupole matrices can be expressed in terms of products of orbital matrices
$L_{i}L_{j}$, e.g. $Q_{zz} = L_{z}^{2}-2/3I$.

For the more symmetric AIM with Dworin-Narath interaction,
the corresponding Kondo Hamiltonian reads
\begin{eqnarray}
	\label{Hkdn0}
	H_{K}^{DN} = J_{p} N_{f}+ J_{s}  \mathbf{S} \cdot
	\mathbf{s} + J_{t}  \mathbf{T} \cdot \mathbf{t}
	+\nonumber\\ J_{ts} \mathbf{(T \otimes
	S)}\cdot( \mathbf{t}\otimes \mathbf{s}).
\end{eqnarray}
In this expression, $ \mathbf{s}$ is the total bath spin operator and
$ t^{\alpha} = \sum_{mm'\sigma}c_{m\sigma}^{\dag}\tau^{\alpha}_{mm'}c_{m'\sigma}$,
$\tau^{\alpha}$ are the Gell-Mann matrices.
$\mathbf{S}$ and $\mathbf{T}$
are the generators of spin-1 representation of SU(2) and the
fundamental representation of SU(3).

\begin{figure*}[t]
	\includegraphics[width=\columnwidth]{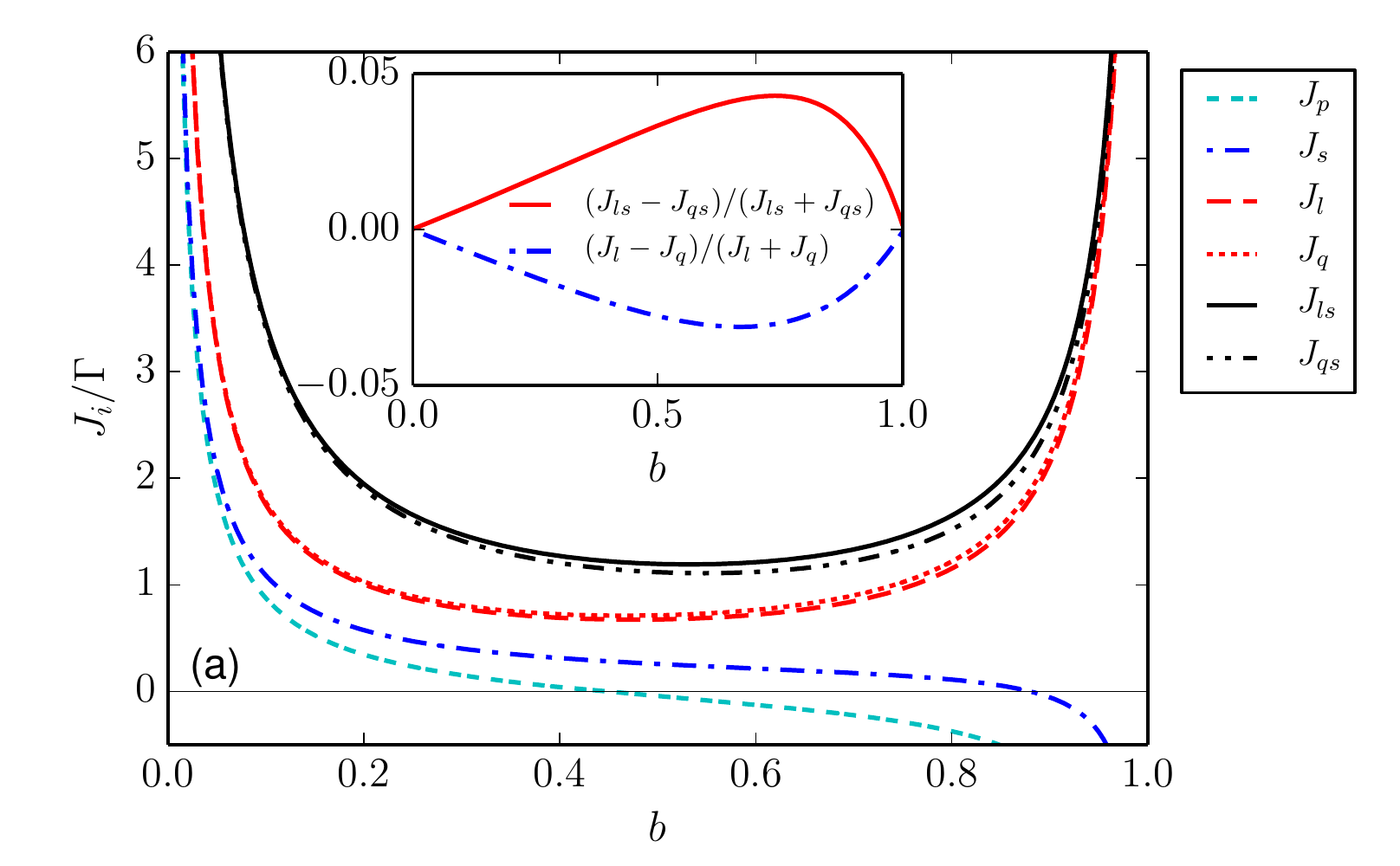}
	\includegraphics[width=\columnwidth]{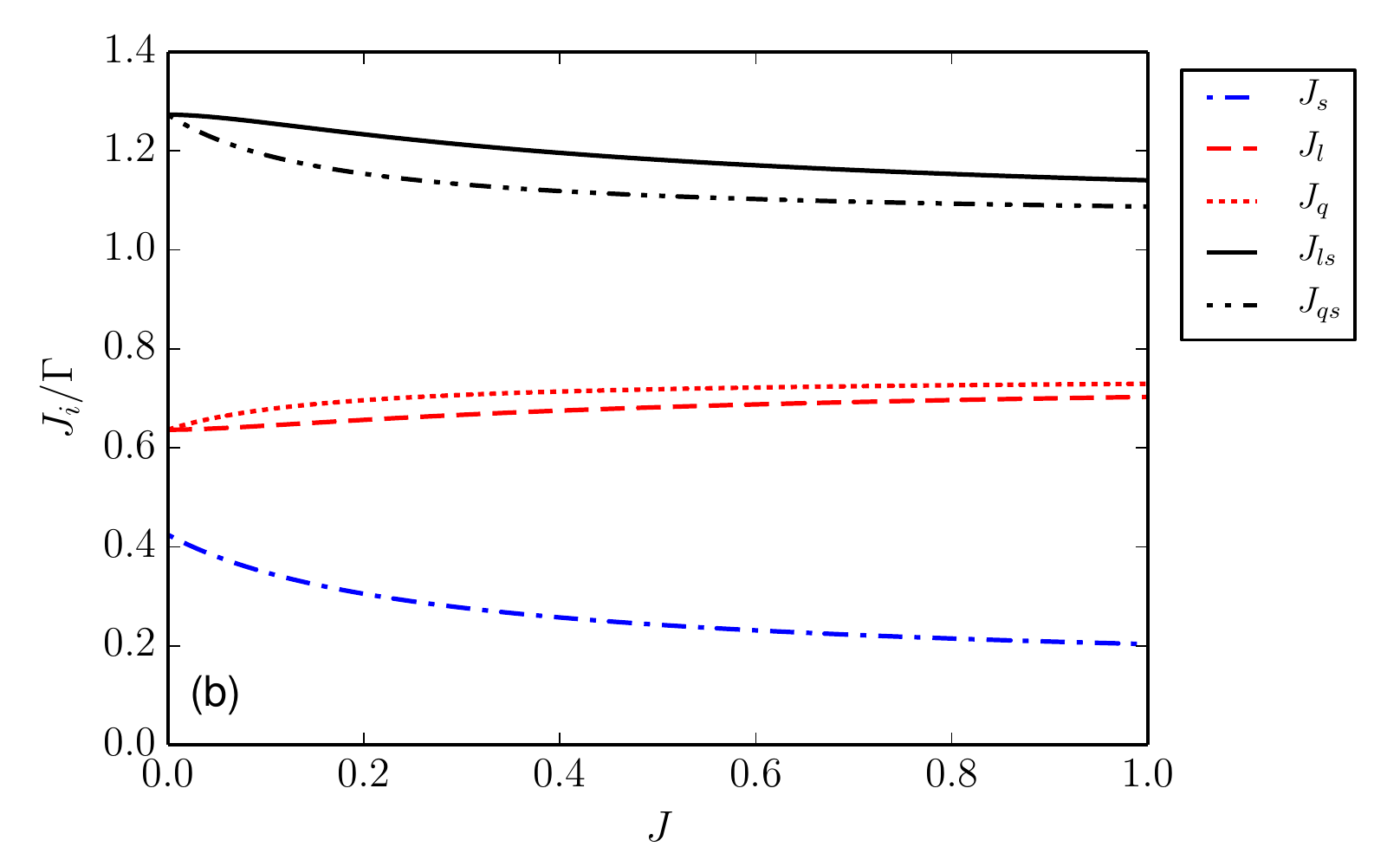}
        \caption{ (a) Bare Kondo exchange coupling constants of the
	effective Kondo-Kanamori
model as a function of the impurity level parameter \( b \).
Model parameters are  $U=3.2, J=0.4$.
The inset shows the relative size of the splittings through ratios $(J_l-J_q)/(J_l+J_q)$
and $(J_{ls}-J_{qs})/(J_{ls}+J_{qs})$.
(b) Bare Kondo couplings as a
function of Hund's coupling $J$ for constant $U_\mathrm{eff}=U-3J = 2$, b=0.5.}
	\label{fig:coupling}
\end{figure*}

The total set of eight generators $\{L,Q\}$ is, in fact, equivalent to
the set of SU(3) generators $\{ T\}$: both sets constitute a basis for
traceless Hermitian $3\times 3$ matrices.
Reducing the SU(3)
orbital symmetry to SO(3) symmetry leads to a splitting of the
orbit-orbit and orbit-quadrupole coupling constants, i.e.,
$J_{t}\rightarrow J_{l}, J_{q}$ and $J_{ts} \rightarrow J_{ls},
J_{qs}$. One of the goals of this work is to study the consequences of
this splitting.

\subsection{Kondo coupling constants}

We calculate the coupling constants for the ground state multiplet with two
electrons occupying the impurity ($N_{d}=2$) that has angular momenta $L=1,S=1$.  In the zero bandwidth limit, $V\rightarrow 0$,
the impurity energy level which determines the impurity occupancy reads:
\begin{equation}
	\epsilon_{0} = \frac{3+2\alpha}{2}J - (1+b) \left[U-J(4-
	\alpha)\right].
\end{equation}
It is measured from the Fermi level. The parameter \( b \in [0,1] \) controls the occupancy
of the impurity before the projection to the $N_d=2$ subspace and
determines the potential
scattering term of the Kondo Hamiltonian. The term is written so that
when $
b\rightarrow 0$ and $b \rightarrow 1$ the atomic $N_d=2$ ground state becomes
degenerate with the atomic lowest states with occupancies $N_d=1$ and $N_d=3$,
respectively.  The excitation energies $\Delta E_{1,3}$ to states with impurity occupancy
$N_{d}\pm 1=1,3$ are presented in Table~\ref{table:1}.
Superscripts $a,b,c$ denote the three
multiplets with charge $N_{d}=3$  having different values of spin and orbital
moment.

\begin{table}[h]
\caption{\label{table:1} Excitation energies. }
    \begin{tabular}{c|c|c|c|c}
   Index & $N_{d}$ & $L$ & $S$   & $\Delta E $\\\hline
    $\Delta E_{1}$ &1 & 1 & 1/2 &      $b (U-(4- \alpha) J)$ \\
   $\Delta E_{3}^{a}$ &3 & 0 & 3/2 & $(1-b) (U-(4-\alpha) J)$ \\
   $\Delta E_{3}^{b}$ &3 & 2 & 1/2 & $(1-b) U+J (b (4-\alpha)+2 (1-\alpha))$ \\
   $\Delta E_{3}^{c}$ &3 & 1 & 1/2 & $(1-b) U+J (b (4-\alpha)+2)$ \\
    \end{tabular}
\end{table}

We note in passing that under the particle-hole
transformation~\cite{Fetter2003} not only the potential scattering
term but also the spin-orbital coupling and the quadrupole-quadrupole
coupling terms of the Hamiltonian in Eq.~\eqref{Hkk0} are odd. As a
result, the two-fold hypercharge degeneracy discussed in
Ref.~\cite{leo04thesis} does not apply even in the absence of
potential scattering.

Next we calculate the  Kondo coupling constants by comparing matrix elements of Hamiltonians
in equations~\eqref{SWtransform} and~\eqref{Hkk0}:
\begin{eqnarray}
	\label{eq:coupling0}
	J_{p} &=&  \frac{V^{2}}{18} \left( \frac{6}{\Delta E_{1}}-\frac{4}{\Delta E_{3}^a}-\frac{5}{\Delta E_{3}^b}-\frac{3}{\Delta E_{3}^c}\right),\\
	J_{s} &=&  \frac{V^{2}}{18} \left( \frac{6}{\Delta E_{1}}-\frac{2}{\Delta E_{3}^a}+\frac{5}{\Delta E_{3}^b}+\frac{3}{\Delta E_{3}^c}\right),\\
	J_{l} &=&  \frac{V^{2}}{12} \left( \frac{6}{\Delta E_{1}}+\frac{8}{\Delta E_{3}^a}-\frac{5}{\Delta E_{3}^b}+\frac{3}{\Delta E_{3}^c}\right),\\
	J_{q} &=&  \frac{V^{2}}{12} \left( \frac{6}{\Delta E_{1}}+\frac{8}{\Delta E_{3}^a}+\frac{1}{\Delta E_{3}^b}-\frac{3}{\Delta E_{3}^c}\right),\\
	J_{ls} &=&  \frac{V^{2}}{6} \left( \frac{6}{\Delta E_{1}}+\frac{4}{\Delta E_{3}^a}+\frac{5}{\Delta E_{3}^b}-\frac{3}{\Delta E_{3}^c}\right),\\
	\label{eq:coupling1}
	J_{qs} &=&  \frac{V^{2}}{6} \left( \frac{6}{\Delta E_{1}}+\frac{4}{\Delta E_{3}^a}-\frac{1}{\Delta E_{3}^b}+\frac{3}{\Delta E_{3}^c} \right).
\end{eqnarray}
These bare Kondo couplings are presented in
Fig.~\ref{fig:coupling}(a) for different values of the parameter $b$.
The spin-spin coupling $J_s$ is substantially smaller than others for
most values of $b$ and changes sign on approaching $b=1$ that
corresponds to the regime of valence fluctuations between $N_d=2$ and
$N_d=3$ (at the degeneracy point between $N_d=2$ and $N_d=3$, the atomic
average occupancy is $30/13\approx 2.3$, while at the degeneracy point
between $N_d=2$ and $N_d=1$, the atomic average occupancy is
$8/5=1.6$). All couplings diverge on approaching the end points $b=0$
and $b=1$ where the cost
for the charge excitations vanishes.  The Kondo model and the
derived couplings for the $N_d=2$, $L=1$, $S=1$ atomic ground state
configuration cease to be valid there.

The results are qualitatively similar to those found for the Dworin-Narath
model in Refs.~\onlinecite{yin2012,Aron2015} with the distinction that for the
Kanamori Hamiltonian the orbital and quadrupole couplings are split:
\[
\begin{split}
J_{q}-J_{l} &= \Delta J/2, \\
J_{qs}-J_{ls} &= -\Delta J
\end{split}
\]
with
\[
	\Delta J = V^{2} \left( \frac{1}{\Delta
	E_{3}^{c}}-\frac{1}{\Delta E_{3}^{b}} \right)=\frac{2J \alpha V^{2}}{\Delta E_{3}^{b}\Delta E_{3}^{c}}.
\]
This results from the different energies of the $L=1$ and $L=2$
three-electron spin-doublet multiplets, caused by the $-\alpha
(J/2)\mathbf{L}^2$ term in the Hamiltonian. For the Kanamori model with $\alpha=1$ the
splitting is largest when the Hund's coupling reaches
\begin{equation}
	J=   \frac{ (1-b) U}{\sqrt{9 b^2+6 b}}.
\end{equation}
For two electrons at the impurity ($b\approx 1/2$) this occurs for \(
J= 0.22U\).
Fig.~\ref{fig:coupling}(b) shows how the splitting
develops as the Hund's coupling $J$ is increased from zero, while
keeping the parameter that controls the charge fluctuations,
$U_\mathrm{eff}=U-3J$, constant. In other words, as $J$ is varied, the
Hubbard repulsion $U$ is adjusted so that the effective impurity
repulsion $U_\mathrm{eff}=E(3)+E(1)-2E(2)=U-3J$ is kept fixed; here
$E(N)$ denotes the energy of the lowest multiplet with occupancy $N$.
The splittings of Kondo couplings are initially linear in $J$, but then
slowly fall off as \( 1/J \).

\subsection{Poor man's scaling analysis}
\label{sec:poor}

\begin{figure*}[!ht]
	\includegraphics[width=2\columnwidth]{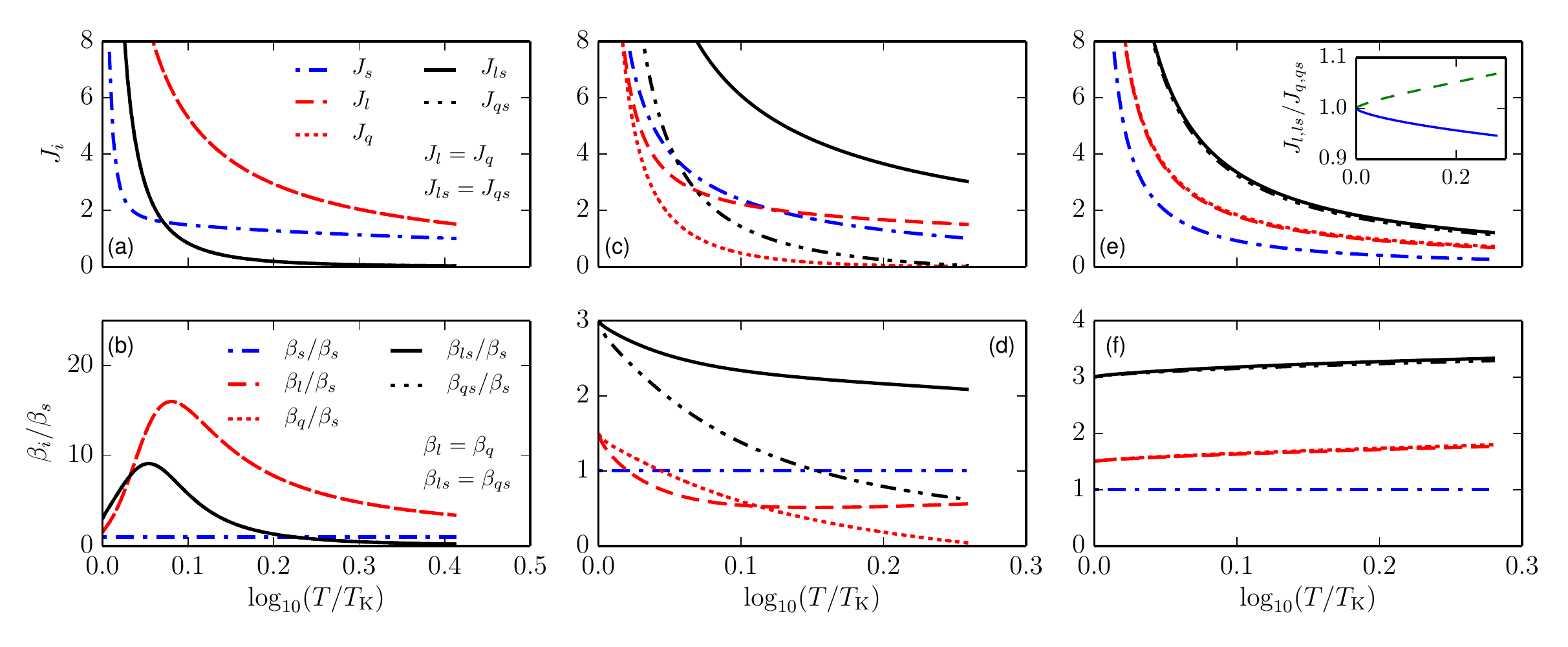}
	\caption{Scaling of the Kondo coupling constants (top) and
	\(\beta\) divided by the spin scaling function \(\beta_s\) (bottom)
	for different initial couplings.
	In (a) and (b) the mixed bare couplings  $J_{ls,qs}= J_{s,q}/100$ are suppressed
	and $J_{l}=J_{q}, J_{ls}=J_{qs}$.
	In (c) and (d) the quadrupole couplings are suppressed,  $J_{q,qs}= J_{l,ls}/100$.
	(e) and (f) the bare couplings that correspond to the
	Anderson impurity model with parameters $U=3.2, J=0.4, \Gamma=1$.
	Inset in (f) shows ratio between the orbital and quadrupole couplings.
	Full and dashed lines correspond to ratios $J_{l}/J_{q}$ and $J_{ls}/J_{qs}$,
	respectively.
	 }
\label{figrgflow1}
\end{figure*}

We now discuss the low-energy physics of the derived Kondo-Kanamori
Hamiltonian within the weak-coupling RG approach.~\cite{anderson1970}
The scaling functions \(\beta_{i}= \mathrm{d}J_{i}/ \mathrm{d}\ln(D)\)
describe the renormalization of the coupling constants as the
half-bandwidth $D$ is progressively reduced. To the lowest order they
read:
\begin{eqnarray}
	\label{eq:couplingA}
	\beta_{p} &=& 0,\\
	\label{eq:scalingS}
	\beta_{s} &=&-\frac{1}{9} \left(3 {J_{ls}}^2+5 {J_{qs}}^2+9 {J_{s}}^2\right), \\
	\label{eq:scalingL}
	\beta_{l} &=&-\frac{1}{16} \left(4 {J_{l}}^2+3 {J_{ls}}^2+5 \left(4 {J_{q}}^2+3{J_{qs}}^2\right)\right), \\
	\beta_{q} &=&-\frac{3}{8} (4 {J_{l}} {J_{q}}+3 {J_{ls}} {J_{qs}}),\\
	\beta_{ls} &=&-\frac{1}{6} (3 {J_{l}} {J_{ls}}+5 {J_{ls}} {J_{qs}}+12 {J_{ls}} {J_{s}}+15 {J_{q}} {J_{qs}}), \\
	\label{eq:couplingB}
	\beta_{qs} &=&-\frac{1}{12} ({J_{qs}} (18 {J_{l}}+7 {J_{qs}}+24 {J_{s}}) \\\nonumber& &+3 {J_{ls}}^2+18 {J_{ls}} {J_{q}}).
\end{eqnarray}
For a particle-hole symmetric band (as is the case for the flat density of
states, $\rho_0=1/2$, used here) the potential scattering operator is marginal,
$\beta_p=0$.

The symmetry of the Hamiltonian is reflected in the scaling
equations. For instance, for vanishing Hund's orbital coupling in the
AIM, the initial orbital and quadrupole coupling constants in
Eq.~\eqref{eq:coupling0} are equal $J_q=J_l$ and $J_{qs}=J_{ls}$. For
such SU(3) orbitally symmetric choice of bare coupling constants, the
respective scaling functions coincide: $ \beta_{q}= \beta_{l},
\beta_{qs}= \beta_{ls}$ and hence $J_q=J_l$ and $J_{qs}=J_{ls}$ also
after RG scaling.

It is interesting to omit the cross-terms by setting
$J_\mathrm{ls}=J_\mathrm{qs}$ which is preserved also after RG flow.
Hence, the spin and orbit coupling constants undergo a separate
scaling in this case.  From the ratio of the two scaling functions:
$\beta_l/\beta_s=3/2 J_l^2/J_s^2$ one sees that besides the larger
bare value of $J_l$ additional factor $3/2$ (the ratio of the orbital
and spin degeneracy) helps the faster renormalization of orbital
couplings. This behavior, associated with the larger SU(3) symmetry
holds only in the case of $J_q=J_l$.

When Hund's coupling is zero, $J=0$, the symmetry is enlarged further
and the model becomes the Coqblin-Schrieffer (CS) model with SU(6)
symmetry. In this case, all the coupling constants are simply related
to each other and can be expressed in terms of a single constant
$J_{CS}=3J_{s} = 2J_{l} = J_{ls}$. The simple relation holds also for
scaling functions $3\beta_{s} = 2\beta_{l} = \beta_{ls}$. (The integer
factors could also be absorbed in the definition of the coupling
constants. See Appendix~\ref{app:symmetric}.)

We numerically solved the scaling equations for three characteristic
cases. We display the results in Fig.~\ref{figrgflow1}. The top panels
show the Kondo couplings and the bottom panels the scaling functions
\(\beta\) divided by the spin scaling function \(\beta_s\).

In left-most panels (a,b) we set the initial values to those of the
SU(6) symmetric case $3J_{s}=2J_{l}=2J_{q}=J_{CS}=1$ but we suppressed
the cross-terms and set $J_{ls}= J_{qs}=J_{CS}/100$. This illustrates
nicely the slower running of the spin coupling that holds until the
cross-terms become large. From this point on, the values of constants
and hence the flow approach that of the Coqblin-Scrieffer SU(6)
symmetric case.

Middle panels (c,d) display the effects of the splitting between the
orbit and quadrupole terms. We used the Coqblin-Schrieffer values
$3J_{s}=2J_{l}=J_{ls}=J_{CS}=1$ for all but the quadrupole and
spin-quadrupole coupling constants that we suppressed
$J_{q}=J_{l}/100, J_{qs}=J_{ls}/100$.  When the quadrupole terms are
small they can be neglected from the scaling equations
Eq.~\eqref{eq:scalingS}-\eqref{eq:couplingB}. In this case initially
the scaling of the spin coupling is faster than the scaling of the
orbit coupling, because the normally large contribution of the
quadrupole terms \( J_{q,qs} \) to \(\beta_{l}\) is not present.
Only when \( J_{q,qs} \) become comparable to \( J_{l,ls} \), the
renormalization of the orbital coupling becomes faster than the
renormalization of the spin coupling and the ratio
$\beta_l/\beta_s$ approaches 3/2.  It is important to note that the
splitting between the orbit and quadrupole and spin-orbit and
spin-quadrupole terms disappears at low energies.

This is also seen in panels (e,f) that show the behavior for realistic
set of initial coupling constants corresponding to the Anderson
model (with parameters $U=3.2, J=0.4, \Gamma=1$). One sees that the
already initially weak splitting between orbital and quadrupole terms
disappears on approaching the low energies (best seen in inset to (e)
that displays the ratio of the two).  Thus the multiplet splitting due
to orbital interaction in the Anderson model becomes insignificant at
low energies.  The SU(3) and SO(3) symmetric models describe the same
low-energy physics. Similar dynamical symmetry generation (or
restoration) has been observed in a number of other quantum impurity
models as well
\cite{Kuzmenko:2002dr,Kikoin:2002fr,Borda:2003ik,Kikoin:2006iq,Keller:2013fk}.

\section{Numerical renormalization group results}
\label{sec:nrg}

Using the NRG technique~\cite{ZitkoNRG} we solve the Kanamori, Dworin-Narath
(DN), and the Kondo impurity model. The NRG results validate the qualitative
insights from the poor man's scaling approach discussed above.  The  two-stage
screening behavior with the spin being screened at a temperature that is
significantly lower than that for the orbital moment occurs in all
three models.

We have implemented an NRG code with conserved quantum numbers
$(Q,S,T)$, corresponding to total charge, total spin and total orbital
angular momentum, i.e., using the U(1)$\otimes$SU(2)$\otimes$SO(3)
symmetry. This allows to perform three-orbital calculations even with
modest computation resources.

\subsection{Comparison between Dworin-Narath, Kanamori, and Kondo-Kanamori results}
\begin{figure}
	\includegraphics[width=\columnwidth]{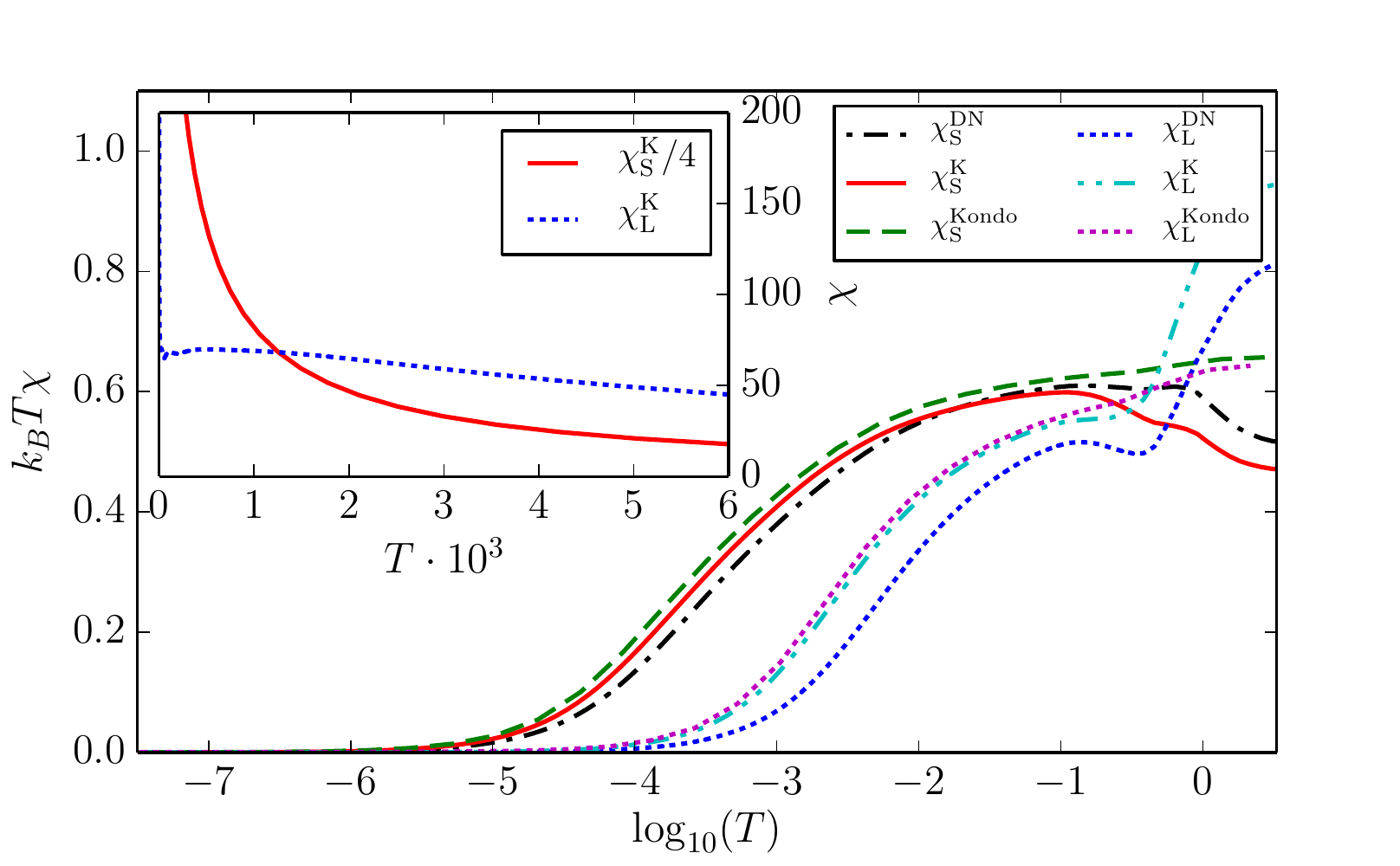}
        \caption{ The NRG results for the effective moments $\chi_{\mathrm{S,L}}
T$ (main panel) and susceptibility $\chi_{\mathrm{S,L}}$ (inset) as a function of temperature for
Dworin-Narath, Kanamori and Kondo-Kanamori models.  $U=3.2, J=0.4, \Gamma=0.1$.
}
	\label{fig:nrg}
\end{figure}

In Fig.~\ref{fig:nrg} we present the temperature dependence of the
effective spin and orbital moments, $\chi_S T$ and $\chi_L T$, where
$\chi_{L,S}$ are the impurity orbital and spin susceptibilities. The
Kanamori results are compared to those for the Kondo model with exchange
couplings set by Eqs.~\eqref{eq:couplingA}-\eqref{eq:couplingB} and
those for the more symmetric Dworin-Narath model.
At high temperatures, the results for different models
significantly differ
due to different high-energy physics. Nevertheless, at lower temperatures
the different models behave alike.  In particular,
the Kondo-Kanamori curves are close to the Kanamori ones (the
differences become
even smaller if the ratio of the interaction to the hybridization is diminished)
which validates our analytical approach.  The Dworin-Narath model behaves similarly,
the main distinction being noticeably higher screening temperature of the orbital
moments.

In the inset to Fig.~\ref{fig:nrg} we present the spin and orbital
susceptibilities. The former is scaled by \( 1/4 \) for easier comparison. The
spin susceptibility is much larger than the orbital susceptibility and the
latter saturates at higher temperatures. This again shows faster screening of
orbital degrees of freedom.  The orbital susceptibility has a weak maximum
before saturating to the low-temperature value. Similar behavior was found in
earlier work~\cite{yin2012}.

\begin{figure}
	\includegraphics[width=\columnwidth]{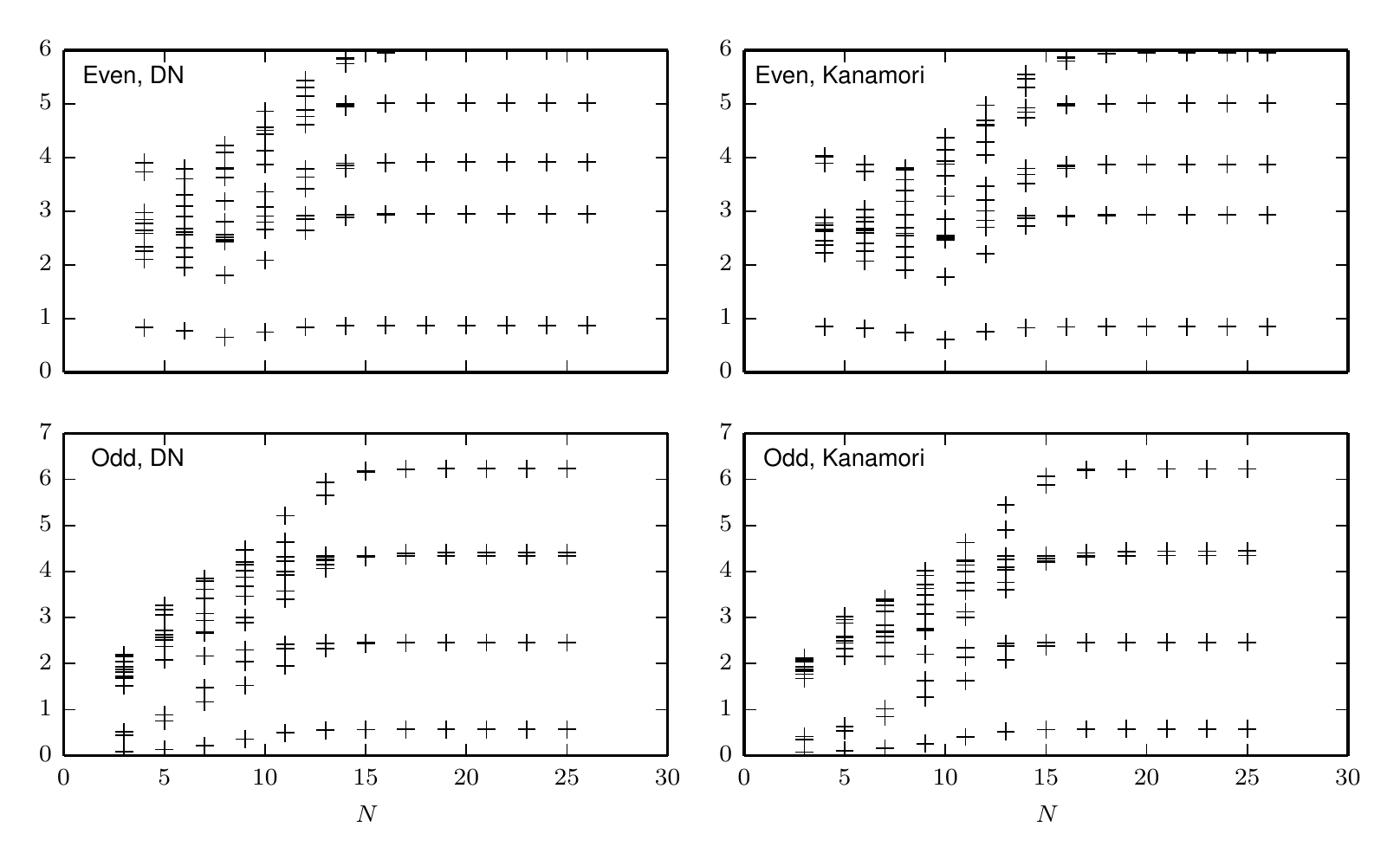}
	\caption{NRG finite size spectra for the DN and Kanamori
	interaction. Parameters are $U=3.2$, $J=0.4$, $N_d=2$.}
	\label{fig:FSS}
\end{figure}

To confirm the asymptotic equivalence of the models, we present in
Fig.~\ref{fig:FSS} the finite size spectra calculated with NRG for the
DN and the Kanamori impurity models as a function of the NRG step. The
two spectra are the same at low energies, which shows that the two
models have the same low-energy Fermi-liquid fixed point with
excitation spectrum parametrized by the quasiparticle phase shift
which is determined by the Friedel sum rule for fixed occupancy
$N_d=2$.

\subsection{Kanamori results at integer occupancy $N_d=2$}

We now discuss the Kanamori model in more detail.
It is convenient to define the spin and orbital Kondo temperatures
as the scale at which the respective effective moment
diminishes below a constant. We take the constant to be 0.07 for spin
and $0.07 l(l+1)/s(s+1)$ for the orbital effective moment~\cite{hewson_book_1993}.
$l,s$ are orbital moment and spin of electrons.
It is of interest to know how the spin \Tksp\ and orbital \Tkorb\ Kondo
temperatures vary with the parameters of the Hamiltonian. We first discuss the
results at an integer occupancy $N_d=2$.

\begin{figure}
	\includegraphics[width=\columnwidth]{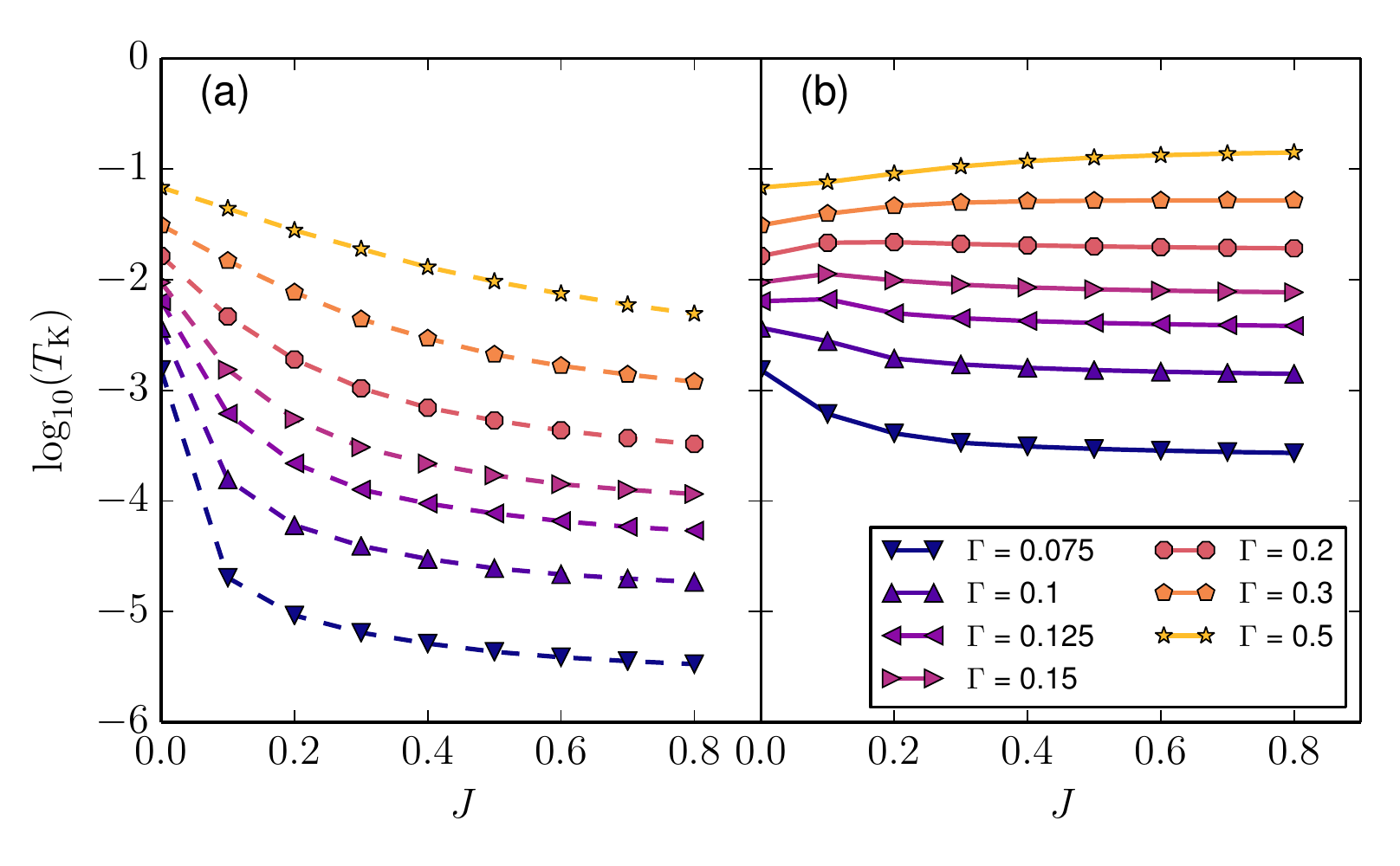}
	\includegraphics[width=\columnwidth]{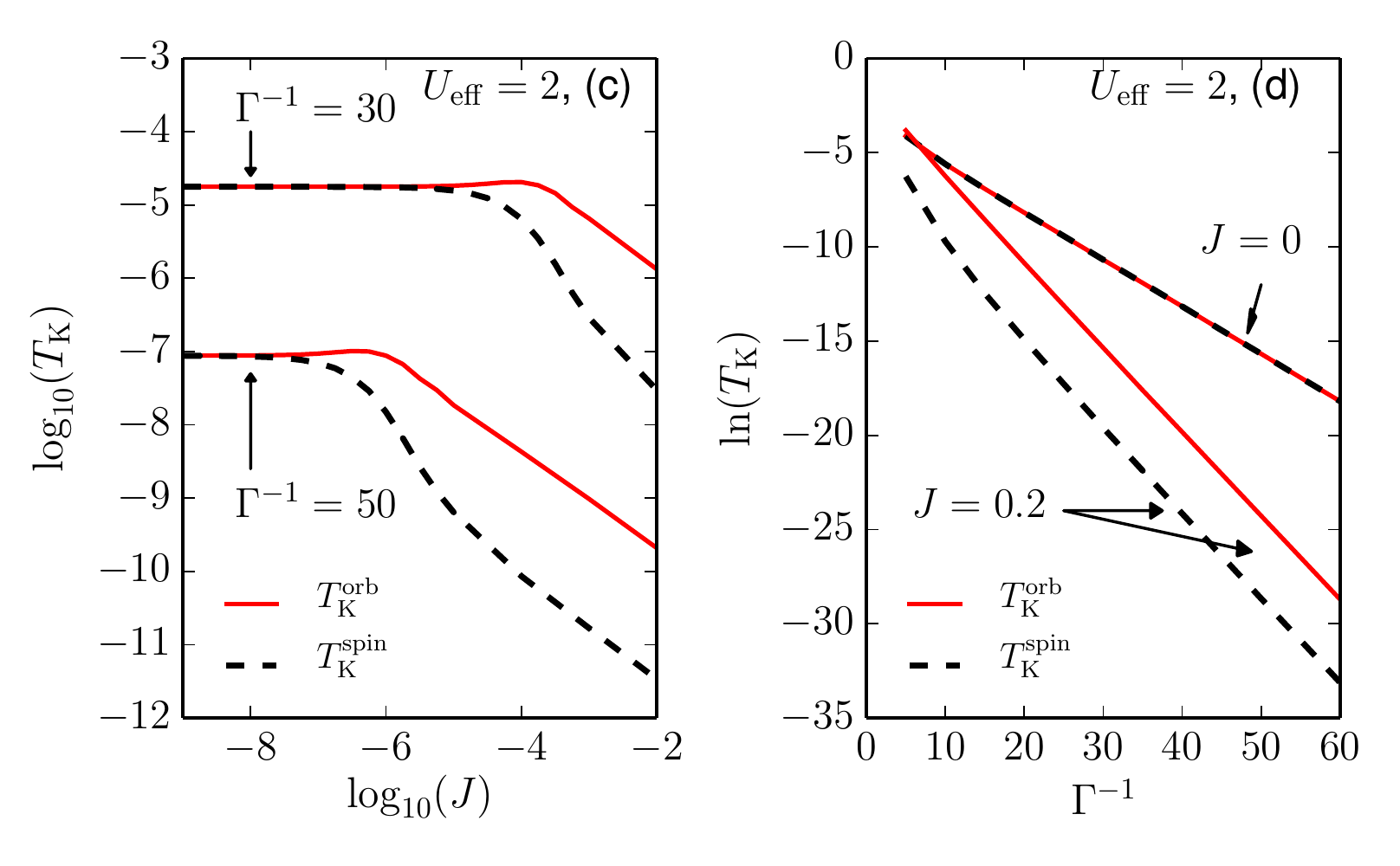}
        \caption{Kondo temperatures for the
         Kanamori model  at fixed $U_{\mathrm{eff}}=2$ and fixed
         impurity occupancy $N_{d}=2$.
    (a) Spin and (b) orbit Kondo
          temperatures as a function of Hund's coupling $J$
          for different values of hybridization
          $\Gamma$. (c) Spin and orbit Kondo temperatures plotted
          versus logarithm of the Hund's coupling $J$. (d) Spin and orbit
          Kondo temperatures as a function of $\Gamma^{-1}$ for
          two values of $J=0, J=0.2$.  }
	\label{fig:Tk_vsJ}
\end{figure}

In Fig.~\ref{fig:Tk_vsJ}(a)--(c) we plot \Tksp\ and \Tkorb\ as a
function of the Hund's rule coupling $J$ for several hybridization
strengths $\Gamma$.  When $J$ is smaller than the Kondo scale of
the $J=0$ model, \( T_{\mathrm{K}}(J=0) = \mTk^{0} \), the moments are
screened before the Hund's coupling has effect.
In this regime symmetry of the model becomes SU(M$\times$N),
hence only a single Kondo scale exists $\mTksp=\mTkorb$.
The Kondo temperature dependence on $J$ is initially slow, but becomes
faster when $J$ becomes larger than \( \mTk^{0} \) as seen from
Fig.~\ref{fig:Tk_vsJ}(c). In addition, close to the $J \sim \mTk^{0}$
point, \Tksp\ becomes smaller than \Tkorb.  Unlike \Tksp\ that
decreases monotonously with $J$, \Tkorb\ has a weak maximum at $J$
above $T_K(J=0)$, which arises as a consequence of an interplay
between the orbital, quadrupole and spin-orbital, spin-quadrupole
interactions. This can be understood from the behavior of the coupling
constants at small $J$.  Namely, upon expanding the Kondo couplings to
first order in $J$ one sees that the orbital-orbital and
quadrupole-quadrupole Kondo interactions increase with $J$, e.g. \(
J_{l}=J_{l}^{0} + \alpha J \), while the other coupling constants
decrease, e.g. \( J_{ls}=J_{ls}^{0} - \beta J \), where \( \alpha,
\beta>0 \) are positive constants.

It is interesting to look at the spin and orbit Kondo temperatures
also as a function of hybridization.  In Fig.~\ref{fig:Tk_vsJ}(d) we
present the logarithms of \Tksp\ and \Tkorb\ as a function of
$\Gamma^{-1}$ for zero and non-zero value of Hund's rule coupling. In
the first case, the spin and orbit Kondo scales are the same
 for all $\Gamma$. Conversely, in the second case, the
spin Kondo temperature is below the orbit Kondo temperature for all
$\Gamma$.  The leading exponential dependence on $\Gamma$ is the same
for both \Tksp\ and \Tkorb, as seen from equal slopes of the lines.
The slopes depend on the repulsion and are $-U_{\mathrm{eff}}/c$ with
(at $N_d=2$) $c \approx 3$ for the zero-$J$ case and $c \approx 4$ for
the finite-$J$ case. The difference is due to increased
degeneracy of multiplets in the $J=0$ case.

\subsection{Kanamori results away from integer filling }

\begin{figure*}[ht]
	\includegraphics[width=2\columnwidth]{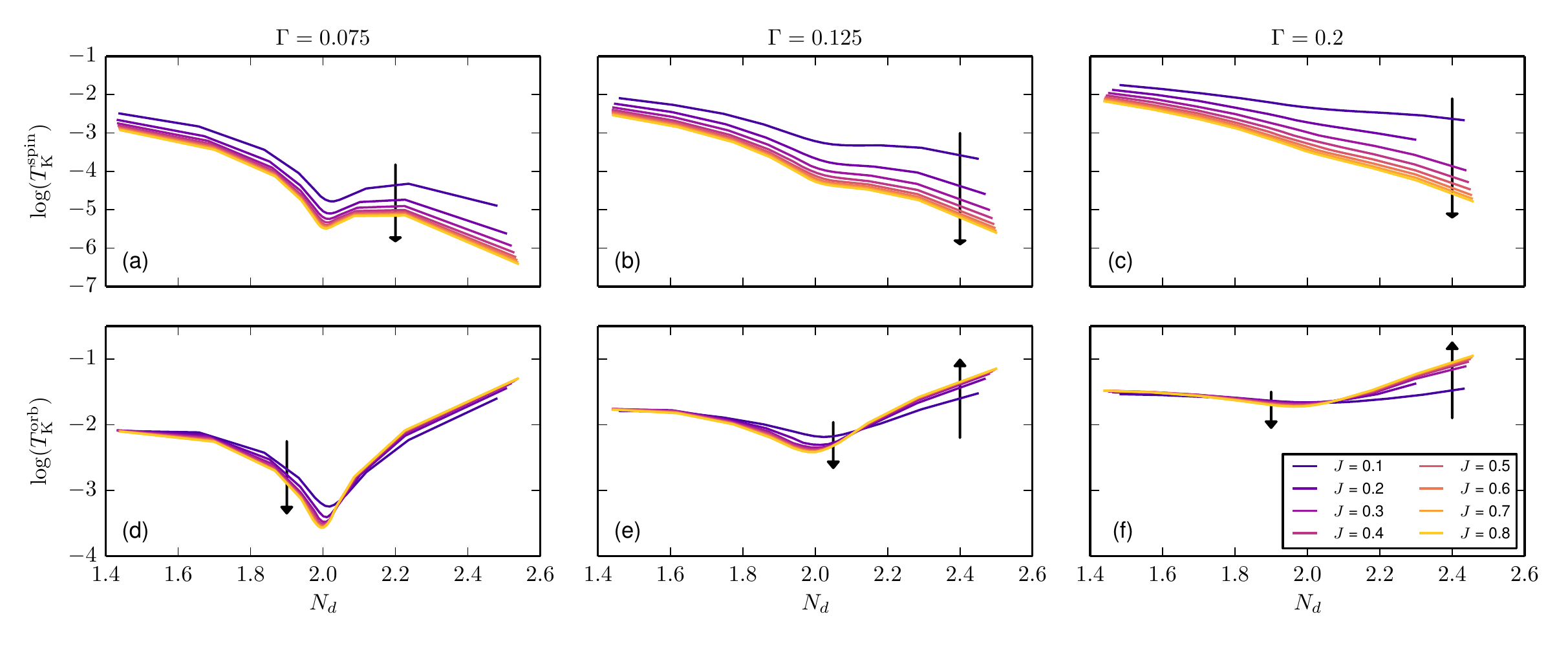}
	\includegraphics[width=0.97\columnwidth]{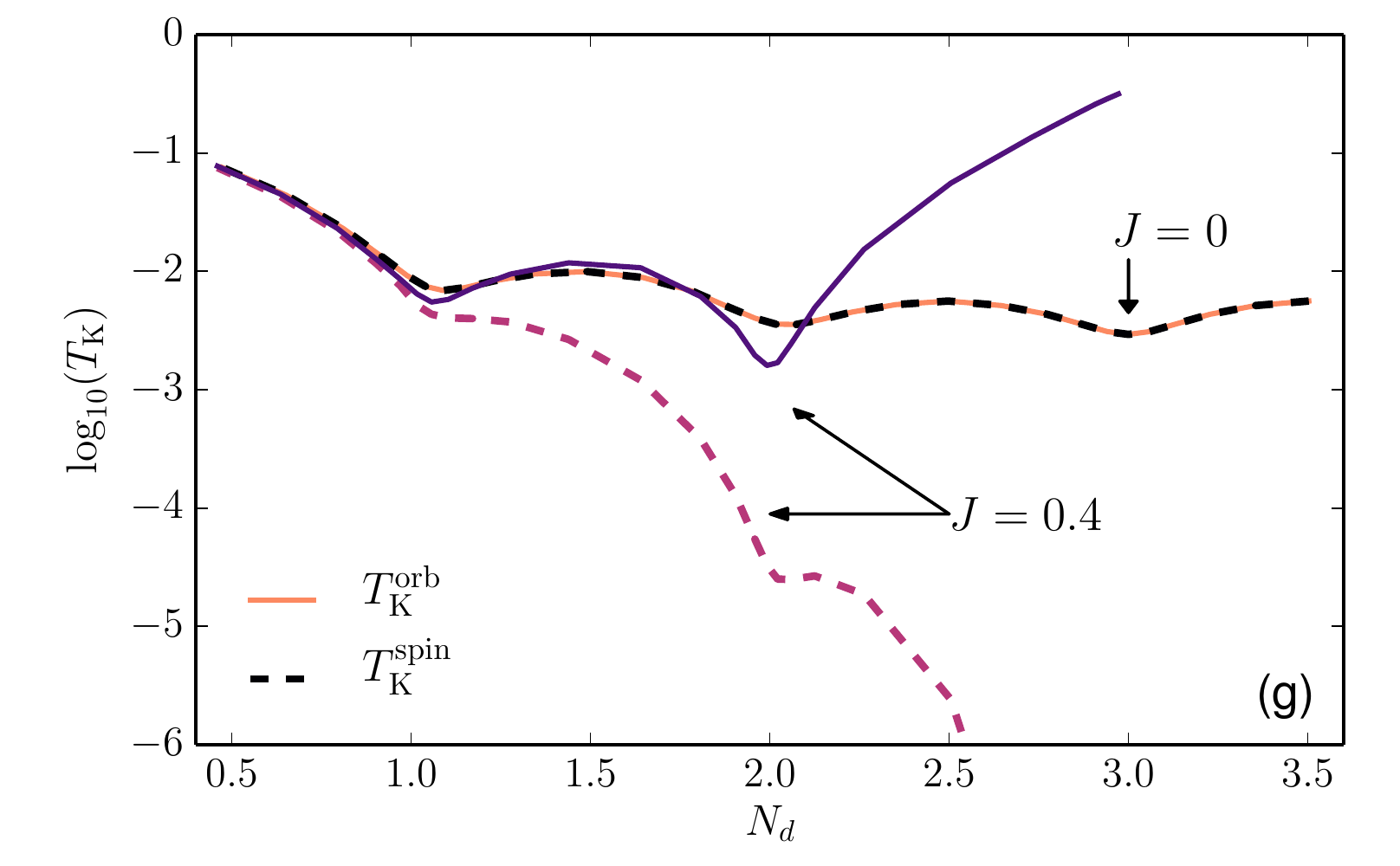}
	\includegraphics[width=0.97\columnwidth]{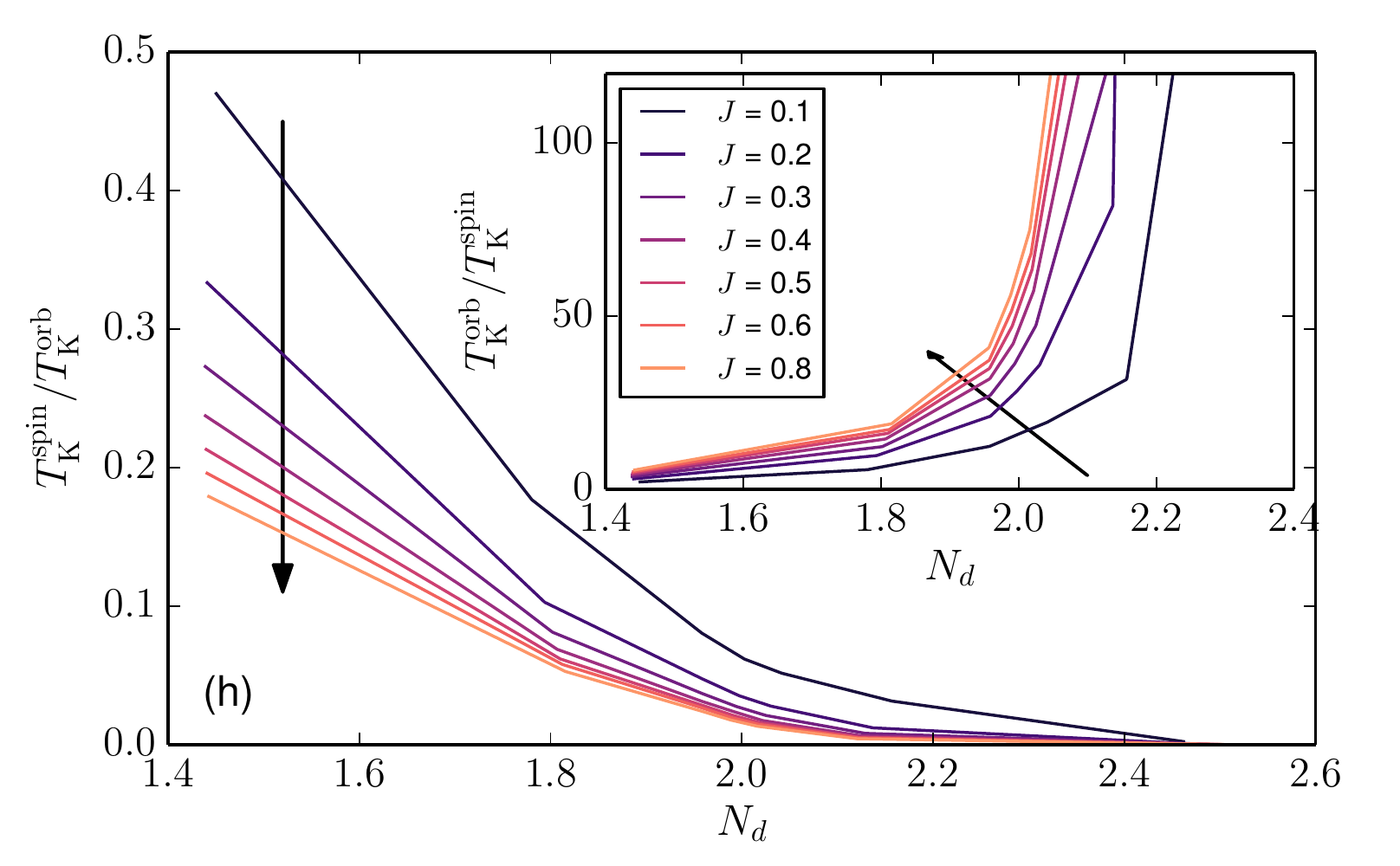}
\caption{ Kanamori model. (a-f) Spin and orbit Kondo temperatures as a
	function of the impurity occupancy $N_d$ for different values
	of the Hund's coupling $J$ at fixed $U_{\mathrm{eff}}=2$,
	(g) Spin and orbit Kondo temperatures in a larger
	region of impurity filling and for zero and non-zero Hund's coupling.
	At $J=0$ the spin and orbit Kondo temperature are the same.
	$\Gamma=0.1$. (h) Ratio between the spin and orbit Kondo
	temperatures. The arrows indicate the direction of
	increasing $J$. }
\label{fig:Tk_vsN}
\end{figure*}

We now turn to the results away from integer filling.  In
Fig.~\ref{fig:Tk_vsN}(a)--(f) we display the Kondo temperatures for
several $\Gamma$ and $J$, still keeping $U_\mathrm{eff}=2$ fixed, as a
function of the impurity occupancy $N_d$ in an interval around 2.
The spin and orbital Kondo temperatures behave differently. \Tksp\
exhibit an overall diminishing trend as $N_d$ is increased towards
half-filling ($N_d=3$) with a shallow minimum at $N_d=2$ that becomes
less pronounced for larger $\Gamma$ where $\log T_K^{\mathrm{spin}}$
is roughly linear in $N_d$.
Conversely, \Tkorb\ increases when occupancy is changed
from $N_d=2$ in both directions for all values of $\Gamma$.

The different behavior of both Kondo temperatures on approaching
half-filling is due to the lowest states at $N_d=3$ having large spin but
vanishing orbital moment, \( L=0, S=3/2 \), thus the screening of the
spin is strongly suppressed because of its large size
\cite{schrieffer_japplphys_1967,nevidomskiy09}, while the orbital
moment is screened at a higher temperature. At half filling,
the notion of orbital Kondo temperature becomes meaningless, as the
orbital moment is zero also in the limit of vanishing hybridization.
This distinction disappears for $J=0$, see Fig.~\ref{fig:Tk_vsN}(g)
where the results for zero and non-zero $J$ are shown in a broader range
of $N_d$. For $J=0$ the spin and orbit Kondo temperatures are the same.

On approaching small occupancies, $N_{d}\lesssim 1$, the Kondo temperatures
rapidly increase and no distinction is seen between zero and non-zero
$J$ cases in panel (g). When there is on average a single
electron in the impurity the Hund's coupling has no effect.

In Fig.~\ref{fig:Tk_vsN}(h) the ratio between the spin and orbital Kondo
temperatures is shown. One sees that $T_K^{\mathrm{orb}}/T_K^{\mathrm{spin}}$
rapidly increases as $N_d$ is increased and at the occupancy $N_d=2$ this ratio
is about 10 and is further increasing as we approach half-filling.

\section{Conclusion}
\label{sec:conc}

We investigated the low-energy behavior of the Kanamori model in the RG
and NRG approaches. We derived the appropriate Kondo model that is
described in terms of spin, orbital, and quadrupole degrees of
freedom. At low energies the splitting between the orbital and
quadrupole coupling constants becomes insignificant, therefore similar
behavior as for a Hamiltonian with a larger SU(3)~\cite{Aron2015}
symmetry can be expected. The NRG results confirm these poor-man's
scaling findings. In particular, both models have the same
strong-coupling Fermi-liquid stable fixed point at low energies and
approach this fixed point in a similar way (in the physically relevant
parameter range). We calculated the dependence of the spin and orbital
Kondo temperatures on interaction parameters, hybridization, and
impurity occupancy. The orbital Kondo temperature is larger, thus
orbital moments are quenched first as the temperature is lowered. This
behavior starts to occur as soon as the Hund's rule coupling is
increased above the Kondo temperature of the problem without the Hund's
rule coupling.  The screening of the spin-moments occurs at a
temperature that is about an order of magnitude smaller~\footnote{The
precise value depends on the parameters. The dominant exponential
dependence on the Coulomb interaction parameters is, however, the same
for the spin and orbital Kondo temperature.}. The ratio of the orbital
Kondo temperature to the spin Kondo temperature becomes particularly
large as the impurity occupancy is increased towards half-filling. Our
results demonstrate that the NRG is capable of treating problems with
realistic three-orbital interactions. This method could hence be used
in the DMFT calculations, too. Another interesting line of
investigation is the analysis of the derived Kondo impurity model for
parameters that do not correspond to the Anderson-type model. Our
preliminary results reveal a rich phase diagram with several distinct
non-Fermi-liquid phases.

\begin{acknowledgments}
We acknowledge the support of the Slovenian Research Agency (ARRS)
under P1-0044.
\end{acknowledgments}

\appendix

\section{Kondo Hamiltonian Derivation}
\label{app:kondo_ham}

In this appendix we derive the Kondo Hamiltonian from the AIM with either Dworin-Narath
or Kanamori interaction using the Schrieffer-Wolff transformation. Kondo
Hamiltonian having SO(3) orbital and SU(2) spin symmetry was earlier written in
terms of unit tensor operators in Ref.~\cite{Hirst1978}.  Kondo Hamiltonian
having SU(M) orbital and SU(N) spin symmetry was derived in
Ref.~\cite{Aron2015}.

The Schrieffer-Wolff transformation reads:
\begin{equation}
    \label{xSWtransform}
    H_{\mathrm{K}} = -P_{n}H_{\mathrm{hyb}} \left( \sum_a \frac{P_{n+1}^{a}}{\Delta
    E_{n+1}^{a}}+\sum_b \frac{P_{n-1}^{b}}{\Delta E_{n-1}^{b}}
    \right)H_{\mathrm{hyb}} P_{n}.
\end{equation}
The projector operator $P_{n}$ projects onto the atomic ground state multiplet with
occupancy \( n=N_{d} \).  The projectors $P^{a}_{n\pm 1}$ project onto
the high-energy atomic multiplets having energy $E^{a}_{n\pm 1}$
(indices $a,b$ denote the different
invariant subspaces with respect to $H_{\mathrm{imp}}$ as presented in the main
text) and the virtual excitation energies are  $\Delta E^{a}_{n\pm 1}=
E^{a}_{n\pm 1}-E_{n}$, $E_{n}$ being the ground-state energy.

We adopt the Einstein summation notation and for the sake of clarity we at first
disregard all the  constants (e.g. $V^{2}/\Delta E$).  The projection operators
to atomic multiplets transform as an identical representation under all symmetry
transformations of the problem, hence the multiplet splitting of the excited states affects
only the coupling constants (we write $\Gamma= H_{\mathrm{hyb}}$):
\begin{equation}
    \sum_{a} \langle n | \Gamma  \frac{P_{n+1}^{a}}{\Delta E_{n+1}^{a}}\Gamma |n\rangle = \sum_{a}
    \frac{1}{\Delta E_{n+1}^{a}} \langle n| \Gamma \Gamma |n \rangle.
\end{equation}
$| n \rangle =P_{n}| \Psi_{ LS} \rangle$ is the ground state with valence $n$, orbital
moment $L$ and spin $S$.  The virtual charge excitation process conserves the
impurity charge, thus $P_{n}d_{j}^{\dagger}d^{\dagger}_{i}P_{n}=0$.
The non-zero
terms in the Kondo Hamiltonian are of the form:
\begin{equation}
    \label{xHid1}
    H_K'=P_{n}\Gamma \Gamma P_{n}=P_{n}(c_{i \sigma_{i}}^{\dagger} d_{i \sigma_{i}}
    d_{k\sigma_{k}}^{\dagger} c_{k \sigma_{k}} + \mathrm{h.c.}) P_{n}
\end{equation}
Next we insert an identity:
\begin{equation}
	\label{xHid}
	c_{i\sigma_{i}}^{\dagger} d_{i \sigma_{i}} d_{k \sigma_{k}}^{\dagger} c_{k\sigma_{k}}=(c_{i \sigma_{i}}^{\dagger}\delta_{i,l}
	\delta_{\sigma_i,\sigma_l}d_{l \sigma_{l}}
	)(d_{k\sigma_{k}}^{\dagger}\delta_{k,j}\delta_{\sigma_k,\sigma_j} c_{j
	\sigma_{j}}),
\end{equation}
and use the following group-theoretical relations~\cite{Cvitanovic1976,varshalovich1989}:
\begin{eqnarray}
	\label{xsuM}
	\delta_{i,l}\delta_{k,j} = \frac{1}{m}\delta_{i,j}\delta_{k,l}+\frac{1}{a} (\tau^b)_{i,j}(\tau^b)_{k,l},\ \mathrm{SU}(m),\\
	\label{xsoM}
	\delta_{i,l}\delta_{k,j} = \delta_{i,k}\delta_{j,l}+\frac{2}{a} (T^b)_{i,j}(T^b)_{k,l},\ \mathrm{SO}(m).
\end{eqnarray}
The generators $\tau, T$ live in the defining (fundamental) representation of
the SU($m$), SO($m$) symmetric Lie group, respectively.  The constant $a$
depends on the normalization of the generators
$\mathrm{Tr}(T^bT^c)=a\delta_{b,c}$ (typically  $a=2$).  In the SU(2) case
$\tau$ are the Pauli matrices and in the SU(3) case $\tau$ are the Gell-Mann matrices.

To  obtain the Kondo Hamiltonian from the AIM with the Dworin-Narath interaction
in terms of spin and orbital operators, we insert the identity~\eqref{xsuM} into
equation~\eqref{xHid} for the spin and orbital degrees of freedom (since both
have SU symmetry).  The relation~\eqref{xsuM} leads to a result in which the
dummy indices associated with the bulk operators $c_{i,j}$ are independent from
the indices associated with the impurity operators, and can be summed over to yield
spin/orbital momentum operators.  The Kondo Hamiltonian with the Dworin-Narath
interaction reads:
\begin{eqnarray}
	\label{xHkdn0}
H_{K}^{DN} = J_{p} N_{f}+ J_{s}  \mathbf{S} \cdot \mathbf{s} +
	J_{t}  \mathbf{T} \cdot \mathbf{t}
	+\nonumber\\ J_{ts} \mathbf{(T \otimes
	S)}\cdot(\mathbf{t}\otimes \mathbf{s}).
\end{eqnarray}
Bath operators are defined as:
\begin{equation}
\begin{split}
\mathbf{s} &= \sum_{m} c^\dag_{m\sigma} \left( \frac{1}{2}
\boldsymbol{\sigma}_{\sigma\sigma'} \right) c_{m\sigma'}, \\
\mathbf{t} &= \sum_\sigma c^\dag_{m\sigma} \boldsymbol{\tau}_{mm'} c_{m'\sigma}.
\end{split}
\end{equation}
$\boldsymbol{\tau},\boldsymbol{\sigma}$ are the Pauli and Gell-Mann matrices, respectively.
$\mathbf{S}$ and $\mathbf{T}$
are the generators of spin-1 representation of SU(2) and the
fundamental representation of SU(3).

On the other hand the relation~\eqref{xsoM}  does not decouple the bulk/impurity
dummy indices due to the term $\delta_{i,k}\delta_{j,l}$.  However, this
problematic term can be, for the  3-dimensional SO(3) symmetric group,
rewritten as
\begin{equation}
	\label{xsoMnew}
	\delta_{i,l} \delta_{k,j}=  \frac{1}{3}\delta_{i,j}
	\delta_{k,l}+\frac{1}{2}T^{c}_{i,j}T^{c}_{k,l}+	 \frac{1}{2}
	Q^{de}_{i,j}Q^{de}_{k,l},
\end{equation}
which does lead to the desired decoupling. Above we used  the
orbital quadrupole operators defined as
\begin{eqnarray}
    \label{xQ}
	Q^{bc}_{i,j} = \frac{1}{2}\left(T^{b}_{i,m}T^{c}_{m,j}+T^{c}_{i,m}T^{b}_{m,j} \right) - \frac{2}{3} \delta_{b,c} \delta_{i,j},\\
	\mathrm{Tr}(Q^{\alpha}Q^{ \beta}) = 2\delta_{ \alpha, \beta},
\end{eqnarray}
which are symmetric and traceless. We derive the
identity~\eqref{xsoMnew} by calculating
$\sum_{b,c}Q^{bc}_{ij}Q^{bc}_{kl}$ and using the
identity~\eqref{xsoM}.  By inserting the identity~\eqref{xsoMnew} for
orbital and~\eqref{xsuM} for spin degrees of freedom into the
Hamiltonian~\eqref{xHid}, we express the Kondo Kanamori Hamiltonian
as:
\begin{eqnarray}
	\label{xHkk0}
	H_{K} = J_{p} N_{f}+ J_{s}  \mathbf{S \cdot s} + J_{l}  \mathbf{L \cdot l}
	+ J_{q}  \mathbf{Q \cdot q} +\nonumber\\ J_{ls} \mathbf{(L \otimes
	S)\cdot(l\otimes s)}+ J_{qs}
	\mathbf{(Q\otimes S)\cdot(q\otimes s)}.
\end{eqnarray}
$ \mathbf{S}, \mathbf{L}, \mathbf{Q}$ ($ \mathbf{s},
\mathbf{l}, \mathbf{q}$) are total impurity (bath) spin, orbit,
orbital-quadrupole  operators respectively.~\cite{louck1984}

\section{RG flow}
\label{app:rg}

In the second order of the
perturbation theory we integrate out the scattering events to
the states close to the band edges, $\pm\epsilon \in  [D- \delta D, D]$.
The first correction to the renormalized Kondo interaction
is
\begin{equation}
\label{SWrg}
	\Delta H_{\mathrm{K}} \approx \frac{1}{\Delta E}
	H_{K} P H_{K}.
\end{equation}
The projector $P$ describes all the scattering events of electrons from
the impurity to the band edges.
The prefactor is \( 1/\Delta E= \rho |\delta D|(E-D+ \epsilon_{k})^{-1}\approx \rho |\delta D| D^{-1}\).
We assume that the conduction band is wide. $D$ is the half-bandwidth, \( E \) is the energy
measured relative to the ground state of the conduction electron gas and can be neglected,
\(\epsilon_{k}\) is the energy of electrons near the Fermi surface and
can also be neglected relative
to \( D \).

In the following we present a convenient way for
calculating the second order corrections to the renormalized Hamiltonian
 using the completeness relations from the previous section.
We will illustrate the procedure on the case of the spin-spin Kondo
interaction term $J \mathbf{S}\cdot \boldsymbol{\sigma}$ for  a single
orbital model with $S=1/2$.
First, we write the impurity operators in terms of the fermionic operators
\begin{equation}
	S^{ \alpha} \rightarrow d_{i}^{\dagger}\sigma_{ij}^{ \alpha}d_{j},
\end{equation}
with additional constraint
\( d_{\uparrow}^{\dagger}d_{\uparrow} + d_{\downarrow}^{\dagger}d_{\downarrow}=1\).
\( d^{\dagger}_{i}, d_{i} \) creates/annihilates an electron
on the impurity with spin \( i \in \{\uparrow, \downarrow\}\), \(\sigma^{\alpha}\) are the Pauli matrices.
The bulk electron spin operator is:
\begin{equation}
	\sigma^{ \alpha} \rightarrow c_{i}^{\dagger}\sigma_{ij}^{\alpha}c_{j},
\end{equation}
\( c^{\dagger}_{i}, c_{i} \) creates/annihilates an electron with spin \( i \)
in the bulk.
The spin-spin operators may be expressed
in terms of Kronecker $\delta$ symbols using the following completeness relation:
\begin{equation}
 	\sum_{ \alpha} (\sigma^\alpha)_{i,j}(\sigma^\alpha)_{k,l} = 2\delta_{i,l}\delta_{k,j} - \delta_{i,j}\delta_{k,l}.
\end{equation}
[For other operators, such as orbital, quadrupole, and mixed operators, one
can derive similar expressions from
Eqs.~\eqref{xsuM},\eqref{xsoM},\eqref{xQ}.]
After inserting the completeness relation we obtain:
\begin{eqnarray}
	 J^{2}
	\sum_{ijkl}(2\delta_{i,l}\delta_{k,j} - \delta_{i,j}\delta_{k,l}) d_{i}^{\dagger}d_{j}c_{k}^{\dagger}c_{l}P \times \\\nonumber
	\times \sum_{mnop}(2\delta_{m,p}\delta_{o,n} - \delta_{m,n}\delta_{o,p}) c_{o}^{\dagger}c_{p}=\\
	=J^{2}\sum_{ijkl}\sum_{mnop} A^{ijkl}_{mnop}P d_{i}^{\dagger}d_{j} d_{m}^{\dagger}d_{n} c_{k}^{\dagger}c_{l}  c_{o}^{\dagger}c_{p}.
\end{eqnarray}
The projector $P$ consists of two contributions:
\begin{equation}
	P = \delta_{jm}( \delta_{lo}+ \delta_{kp}).
\end{equation}
The first term \( \delta_{jm} \) follows from the single-occupancy constraint of auxiliary fermions,  while the second term $\delta_{lo}+ \delta_{kp}$
describes the processes that involve scattering of electrons/holes to the upper/lower
band edge.
In the expressions one can use   \( c^{\dagger}_{\sigma k}c_{\sigma k}=0 \) for the electron states $k$ in the upper band edge that are assumed empty and
 \( c^{\dagger}_{\sigma k }c_{\sigma k}=1 \) for the electron states $k$  at the lower band edge that are assumed filled.

Now we sum over the indices \( m,o \) to eliminate Kronecker
$\delta$ symbols that come from the projection operator.
The contribution of the electron scattering to the upper band edge reads:
\begin{equation}
	J^{2}\sum_{ijkl}\sum_{np} A^{ijkl}_{jnlp} d_{i}^{\dagger}d_{n} c_{k}^{\dagger}c_{p}.
\end{equation}
Next we sum over the dummy indices \( j,l \). The correction to the Kondo exchange reads:
\begin{eqnarray}
\label{b10}
	J^{2}\sum_{ik np} (-4 \delta_{ip} \delta_{kn}+5 \delta_{in} \delta_{kp}) d_{i}^{\dagger}d_{n} c_{k}^{\dagger}c_{p}=\\
	=-2J^{2} \mathbf{S} \cdot \boldsymbol{\sigma}
	+ 3J^{2} \sum_{ik np} \delta_{in} \delta_{kp} d_{i}^{\dagger}d_{n} c_{k}^{\dagger}c_{p}.
\end{eqnarray}
This result has the same form as the initial exchange interaction with
an additional potential scattering term. A contribution from the
scattering to the lower band edge is obtained in a similar fashion;
the exchange term is the same, while the potential scattering term has
an opposite sign and therefore cancels out that in Eq.~\eqref{b10}
since we have assumed a particle-hole symmetric conduction band. We
recover the standard $\beta$ function of the $S=1/2$ Kondo model.

Similar approach can be used to tackle the multi-orbital problem.
The scaling functions for a flat band, general number of orbitals $M$ and $N=2$ are:
\begin{widetext}
\begin{eqnarray}
\beta_{s} = \frac{M \left({J_{ls}}^2-M \left({J_{ls}}^2+2
{J_{s}}^2\right)\right)-{J_{qs}}^2 \left(M^2+M-2\right)}{2 M^2},\\
\beta_{l}=\frac{1}{16} \left(-4 {J_{l}}^2 (M-2)-3 {J_{ls}}^2 (M-2)-(M+2) \left(4
{J_{q}}^2+3 {J_{qs}}^2\right)\right),\\
 \beta_{q} =-\frac{1}{8} M (4 {J_{l}}{J_{q}}+3 {J_{ls}} {J_{qs}}),\\
  \beta_{ls} =-\frac{{J_{ls}} \left(M ({J_{l}} (M-2)+4 {J_{s}})+{J_{qs}} \left(M^2-4\right)\right)+{J_{q}} {J_{qs}} M (M+2)}{2M},
\\ \beta_{qs} =-\frac{2 {J_{qs}} M ({J_{l}} M+4 {J_{s}})+{J_{ls}} M({J_{ls}} (M-2)+2 {J_{q}} M)+{J_{qs}}^2 \left(M^2+2 M-8\right)}{4 M}.
\end{eqnarray}
\end{widetext}
When $ \alpha=0, J_{q}=J_{l}, J_{qs}=J_{ls}$ and
results are the same as obtained in Ref.~\cite{Kuramoto1998, Aron2015} for the model
with SU(M) orbital symmetry.

\section{Rescaled Kondo Hamiltonian}
\label{app:symmetric}

In the Coqblin-Schrieffer model the coupling constants are related to each
other: $3J_{p,s}=2J_{l,q}=J_{ls,qs}$.
We introduce rescaled coupling constants:
$\tilde J_{p,s}=3J_{p,s}, \tilde J_{l,q}=2J_{l,q}, \tilde J_{ls,qs}=J_{ls,qs}$.
The Kondo Hamiltonian in terms of rescaled couplings reads:
\begin{eqnarray}
	H_{K} &=& \tilde J_{p}/3 N_{f}+ \tilde J_{s}/3  \mathbf{S \cdot s} + \tilde J_{l}/2  \mathbf{L \cdot l}
	+
	\tilde J_{q}/2  \mathbf{Q \cdot q} +
	\nonumber\\
	& &
	\tilde J_{ls} \mathbf{(L \otimes
	S)\cdot(l\otimes s)}+ \tilde J_{qs}
	\mathbf{(Q\otimes S)\cdot(q\otimes s)}.
\end{eqnarray}
Hence the rescaled Kondo couplings are written in a more symmetric form:
\begin{eqnarray}
	\tilde J_{p} &=&  \frac{V^{2}}{6} \left( \frac{6}{\Delta E_{1}}-\frac{4}{\Delta E_{3}^a}-\frac{5}{\Delta E_{3}^b}-\frac{3}{\Delta E_{3}^c}\right),\\
	\tilde J_{s} &=&  \frac{V^{2}}{6} \left( \frac{6}{\Delta E_{1}}-\frac{2}{\Delta E_{3}^a}+\frac{5}{\Delta E_{3}^b}+\frac{3}{\Delta E_{3}^c}\right),\\
	\tilde J_{l} &=&  \frac{V^{2}}{6} \left( \frac{6}{\Delta E_{1}}+\frac{8}{\Delta E_{3}^a}-\frac{5}{\Delta E_{3}^b}+\frac{3}{\Delta E_{3}^c}\right),\\
	\tilde J_{q} &=&  \frac{V^{2}}{6} \left( \frac{6}{\Delta E_{1}}+\frac{8}{\Delta E_{3}^a}+\frac{1}{\Delta E_{3}^b}-\frac{3}{\Delta E_{3}^c}\right),\\
	\tilde J_{ls} &=&  \frac{V^{2}}{6} \left( \frac{6}{\Delta E_{1}}+\frac{4}{\Delta E_{3}^a}+\frac{5}{\Delta E_{3}^b}-\frac{3}{\Delta E_{3}^c}\right),\\
	\tilde J_{qs} &=&  \frac{V^{2}}{6} \left( \frac{6}{\Delta E_{1}}+\frac{4}{\Delta E_{3}^a}-\frac{1}{\Delta E_{3}^b}+\frac{3}{\Delta E_{3}^c} \right).
\end{eqnarray}
Notice that in the limit of vanishing Hund's coupling $J=0$, $\Delta E_{i}=\Delta E$,
and all the couplings are the same and so are the scaling functions:
\begin{eqnarray}
	\label{eq:couplingAbis}
	\tilde \beta_{p} &=& 0,\\
	\tilde \beta_{s} &=&-\frac{1}{3} \left(3 {\tilde J_{ls}}^2+5 {\tilde  J_{qs}}^2+ {\tilde J_{s}}^2\right), \\
	\tilde \beta_{l} &=&-\frac{1}{8} \left({\tilde J_{l}}^2+3 {\tilde J_{ls}}^2+5 \left( {\tilde J_{q}}^2+3{\tilde J_{qs}}^2\right)\right), \\
	\tilde \beta_{q} &=&-\frac{3}{4} ( {\tilde J_{l}} {\tilde J_{q}}+3 {\tilde J_{ls}} {\tilde J_{qs}}),\\
	\tilde \beta_{ls} &=&-\frac{1}{12} (3 {\tilde J_{l}} {\tilde J_{ls}}+10 {\tilde J_{ls}} {\tilde J_{qs}}+8 {\tilde J_{ls}} {\tilde J_{s}}+
	\\\nonumber & &
	+15 {\tilde J_{q}} {\tilde J_{qs}}), \\
	\tilde \beta_{qs} &=&-\frac{1}{12} ({\tilde J_{qs}} (9 {\tilde J_{l}}+7 {\tilde J_{qs}}+8 {\tilde J_{s}})+ \\\nonumber& &+3 {\tilde J_{ls}}^2+9 {\tilde J_{ls}} {\tilde J_{q}}).
\end{eqnarray}

\section{Comparison between Kanamori and Dworin-Narath models}
\label{app:kan_vs_dn}

\begin{figure*}
	\includegraphics[width=2\columnwidth]{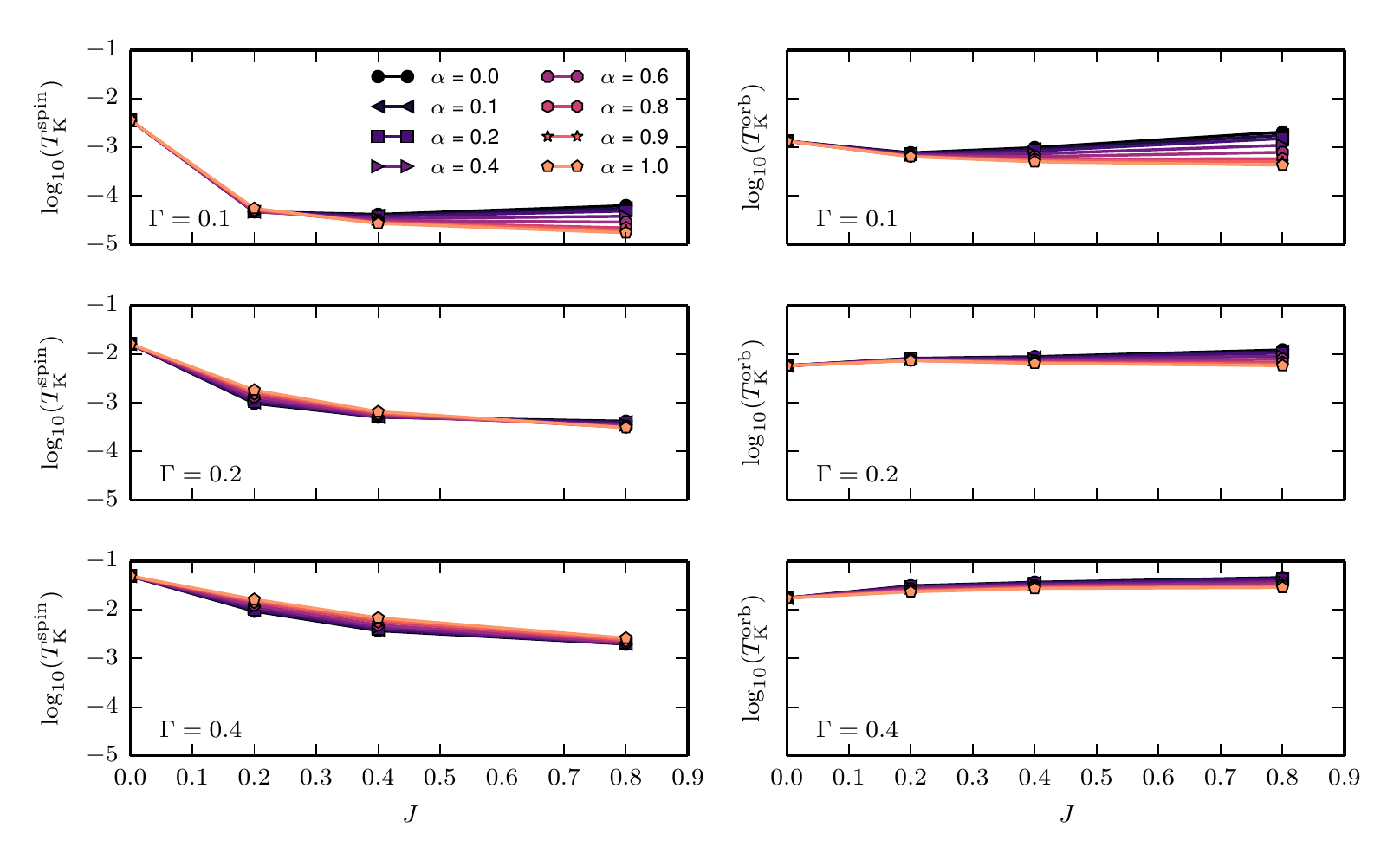}
	\caption{ Spin and orbital Kondo temperatures as a function
	of Hund's coupling $J$ for different values of parameter $ \alpha$ (\(\alpha=0\) DN interaction,
	\(\alpha=1\) Kanamori interaction). Model parameters are $U_\mathrm{eff}=2, N_d=2.$}
	\label{fig:nrg4}
\end{figure*}

Using parameter $\alpha$ (Eq.~\eqref{eq:imp-general} in the main text)
the impurity interaction can be continuously tuned between the
Dworin-Narath ($ \alpha = 0$) and the Kanamori ($ \alpha = 1$) form.
Even though the SO(3) orbital symmetry is dynamically restored to
SU(3) at low energies and hence the behavior of the two models is
similar there are quantitative differences that we illustrate here.

In Fig.~\ref{fig:nrg4} we present the spin and the orbit Kondo
temperatures as a function of Hund's coupling for different values of
\(\alpha\).  Overall a qualitatively similar behavior is found. At
small hybridizations up to an order of magnitude difference is found
for large $J$.  For small hybridization the spin Kondo temperature for
Dworin-Narath is non-monotonic at large $J$ which is not the case for
the Kanamori model. The calculated Kondo temperatures there differ by
an order of magnitude between the two models which can be important
for realistic DMFT calculations where the quantitative agreement with
experiments is desired. Despite the overall similarity of the
Dworin-Narath and Kanamori results, the more realistic Kanamori
interaction needs to be used there.

\end{document}